# Structural-dynamic behavior of histamine in solution: the role of water models

Dmytro A. Gavryushenko[1], N. Atamas [1,2], Oleg K. Myronenko[1]

1. Taras Shevchenko National University of Kyiv, Kyiv, Ukraine
2. BOKU University, Vienna, Austria

*Abstract:* A highly diluted aqueous solution of histamine at T = 300K was studied by molecular dynamics using the TIP3P and SPC/E water models. It was shown that the local structure of the solution around histamine is determined by local Coulomb interactions and hydrogen bonds and is practically independent of the choice of the water model. Dynamic analysis based on the mean square displacement functions revealed a significant dependence of the diffusion behavior of histamine on the water model. It was found that the TIP3P water model leads to overestimated values of the diffusion coefficients of water and histamine and a transition to the diffusion mode of motion. It was found that the SPC/E water model provides slower dynamics of the solution components, and the values of the diffusion coefficients are in better agreement with experimental data. It was shown that the dynamics of histamine is highly sensitive to the choice of the water model, and the SPC/E model is more suitable for the correct description of the dynamic properties of the "histamine–water" system under physiological conditions.

# INTRODUCTION

Imidazole ligands play a key role in coordination chemistry, biochemistry, and pharmacology due to their ability to participate in proton transfer, hydrogen bond formation (HB) and transition metal ion coordination. The central structural element of such compounds is the imidazole ring, a five-membered heterocycle with two nitrogen atoms, one of which exhibits pronounced σ-donor properties. Imidazole moieties are widely represented in the active centers of enzymes and metalloproteins, where they participate in catalytic processes, electron transfer, and stabilization of protein structures [1,2]. The high sensitivity of the imidazole ring to protonation makes such systems particularly dependent on the properties of the aqueous environment. A convenient model molecule for studying the behavior of imidazole ligands in water is histamine $C_5H_9N_3$, which contains an imidazole ring and a primary amino group. The histamine molecule can exist in several protonated forms. Therefore, the dynamics and conformational equilibrium in solution are largely determined by the structure and dynamics of the hydration shell around histamine dissolved in water. Due to these properties, histamine is widely used to study solvation processes, hydrogen bond dynamics, and diffusion characteristics of small biologically active molecules [3]. At the same time, a correct description of the properties of water is a critically important condition for correct molecular dynamics (MD) modeling of ligand-water systems, since water is a complex associated fluid with a wide range of relaxation time scales. Therefore, even relatively small differences in the parameters of water models can lead to significant discrepancies in the reproduction of density, self-diffusion coefficients, dielectric constants, and lifetimes of the HB, which is directly reflected in the results obtained regarding the structural-dynamic behavior of the ligands dissolved in it. Among the classical rigid water models that are widely used in biomolecular MD calculations, a special place is occupied by the three-point models TIP3P and SPC/E [4]. Despite the formal structural similarity, these models differ significantly in reproducing the experimental thermodynamic and dynamic properties of water at the physiologically relevant temperature $T \approx 300$ K. Modern studies show that the choice of the water model directly affects the dynamic properties of ligands in solution [5]. The temperature T=300K is physiologically relevant and optimal for analyzing the differences between water models [6], since it is in this range that the most pronounced differences in the dynamics of the HB nets and the transport properties of the solvent are detected [7]. For histamine, which plays an important role in immune reactions, neurotransmission, and regulation of vascular tone, a correct description of the dynamics of hydration at this temperature is of great importance for understanding its functional activity. Despite the availability of data on the static hydration structure of histamine, the dynamic characteristics of the histamine-water system remain poorly understood, especially in the context of a systematic comparison of different water models. In this context, the study of a dilute histamine solution at T= 300 K using the TIP3P and SPC/E models allows us to isolate the influence of the microscopic properties of the solvent and assess how the choice of the water model affects the dynamics of the imidazole ligand.

**RESEARCH METHODOLOGY**

The molecular dynamics method for studying the structural and dynamic properties of water-histamine systems was implemented using the DL_POLY_4.06 [8] and DL_FIELD [9] software packages. The DL_FIELD package was used to generate the parameters and charges of the histamine molecule, which in the framework of the study was described by OPLS force field parameters (Fig. 1). Interactions between water molecules were modeled using the SPC/E or TIP3P potential. All calculations were performed with a time step of 2 fs for systems consisting of one histamine molecule and ~887 water molecules at a temperature of T = 300K. The following molecular dynamics calculations were performed using the DL_POLY_4.06 software package in a cubic cell with periodic boundary conditions. The cell volume was calculated based on experimental data on the density of water at T = 300K. The initial configuration of the studied water-histamine systems was equilibrated in the NVE ensemble at $1.5 \times 10^6$ steps, after which the final structure was used for further equilibration and calculations in the NVT ensemble with a Berendesen thermostat at $2.0 \times 10^6$ steps. Short-range Coulomb interactions were taken into account by assigning charges to each atom in the molecule, and long-range electrostatic interactions were taken into account by Ewald summation.

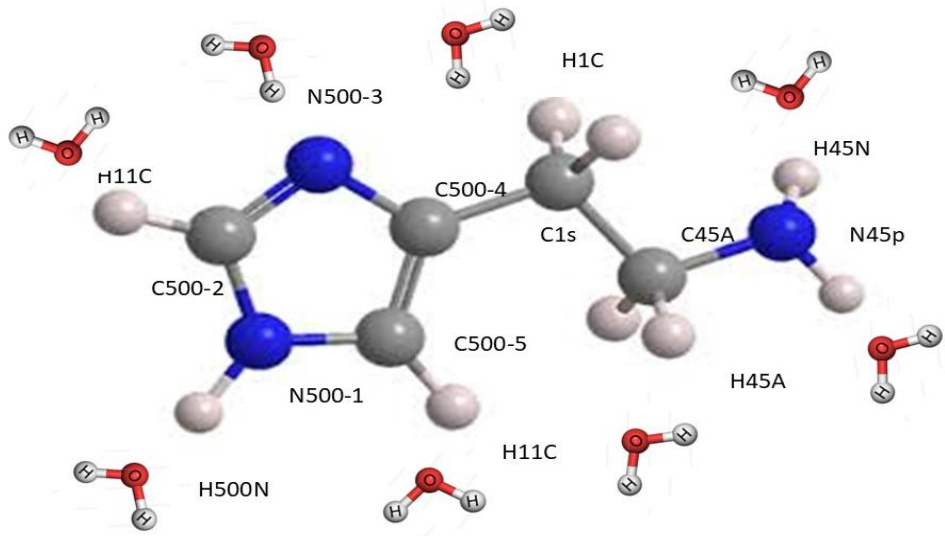

**Fig. 1.** Model representation of histamine (C₅H₉N₃)

In the studied systems, intermolecular interactions in the system were described as the sum of Lennard-Jones and Coulomb interactions [10]:

$$U = U_{L-J} + U_{Coulon} = \sum_{ij} 4\varepsilon_{ij}\left[\left(\frac{\sigma_{ij}}{r_{ij}}\right)^{12} - \left(\frac{\sigma_{ij}}{r_{ij}}\right)^{6}\right] + \sum_{ij} \frac{q_i q_j}{r_{ij}}, \quad (1)$$

where the parameters $\sigma_{ij}$ and $\varepsilon_{ij}$ correspond to the Lennard-Jones potential, which describes the interactions between atoms i and j in the molecule, $r_{ij}$ is the distance between them, and $q_i$ and $q_j$ are the charges of the atoms, respectively. The interaction parameters between the atoms of the histamine molecule and water were calculated using the Lorentz-Bartholo rule.

**RESULTS AND DISCUSSION**

*Local structure of the water-histamine system*

To understand intermolecular interactions in biologically significant systems, such as water–histamine, it is important to assess not only the influence of the solute (histamine) on the formation of the local structure of the solution, but also the role of the chosen water model used in the simulation. For this purpose, the SPC/E and TIP3P water model representations were used to study the structural features of the water–histamine system at a temperature of T=300K. In addition, special attention was paid to determining the influence of the choice of the water model on the specificity of the interaction of water molecules with the atoms of the imidazole ring of histamine. Structural analysis of solutions was carried out based on the analysis of radial distribution functions (RDF), which allow quantitatively determining the structural parameters of solutions (distances between water molecules, distances between water and histamine molecules, estimating the size of the free space between them, determining the probability and characteristics of the HB between histamine and water). In general, the *RDF $G_{XY}(r)$* gives the probability of finding particles of type y near particles of type *x* and can be calculated using the following equation [10]:

$$G_{XY}(r) = \frac{\langle N_y(r, r + dr)\rangle}{4\pi\rho_y r^2 dr} \qquad (2)$$

In this equation, the numerator describes the average number of particles y in the spherical layer (*r, r+dr*), and the denominator normalizes the distribution such that $G_{XY}(r) = 1$ when $N_y = \rho_y$, where $\rho_y$ is the density of the system.

Fig. 1 shows the RDF for water-histamine systems, in which water is represented within the TIP3P (left column) and SPC/E (right column) models. The RDF (Fig. 2) shows that water molecules can be located at a distance of ~3.0 Å from the N45p atom due to the Coulomb interaction between the N45p atom and the $H^w$ hydrogen of the water molecule. In addition, due to the Coulomb interaction between the N45p nitrogen atom and the $O^w$ oxygen atom of the water molecule, water molecules are also observed at a distance of ~2.8 Å in the vicinity of the $NH_2$ group, which leads to a change in the local structure of water in the $NH_2$ group region. According to the RDFs data (Fig. 2), the clearly expressed position of the first RDF minimum for the N500-3 ... $O^W$ and N500-3 ... $H^W$ interactions indicates the formation of the first hydration shell with a size of up to ~2.7 Å. The size of the second hydration shell in the region of the N500-3 atom of the imidazole ring is ~4.5

Å. In addition, RDF shows that in the region of the N500-1 atom of the imidazole ring, water molecules can be located at a distance of ~3.9 to ~4.8 Å. The RDF analysis allows us to establish that several hydration shells are formed in the vicinity of the N500-1 atom, and the radius of the first hydration sphere in the vicinity of this atom is ~5.7 Å.

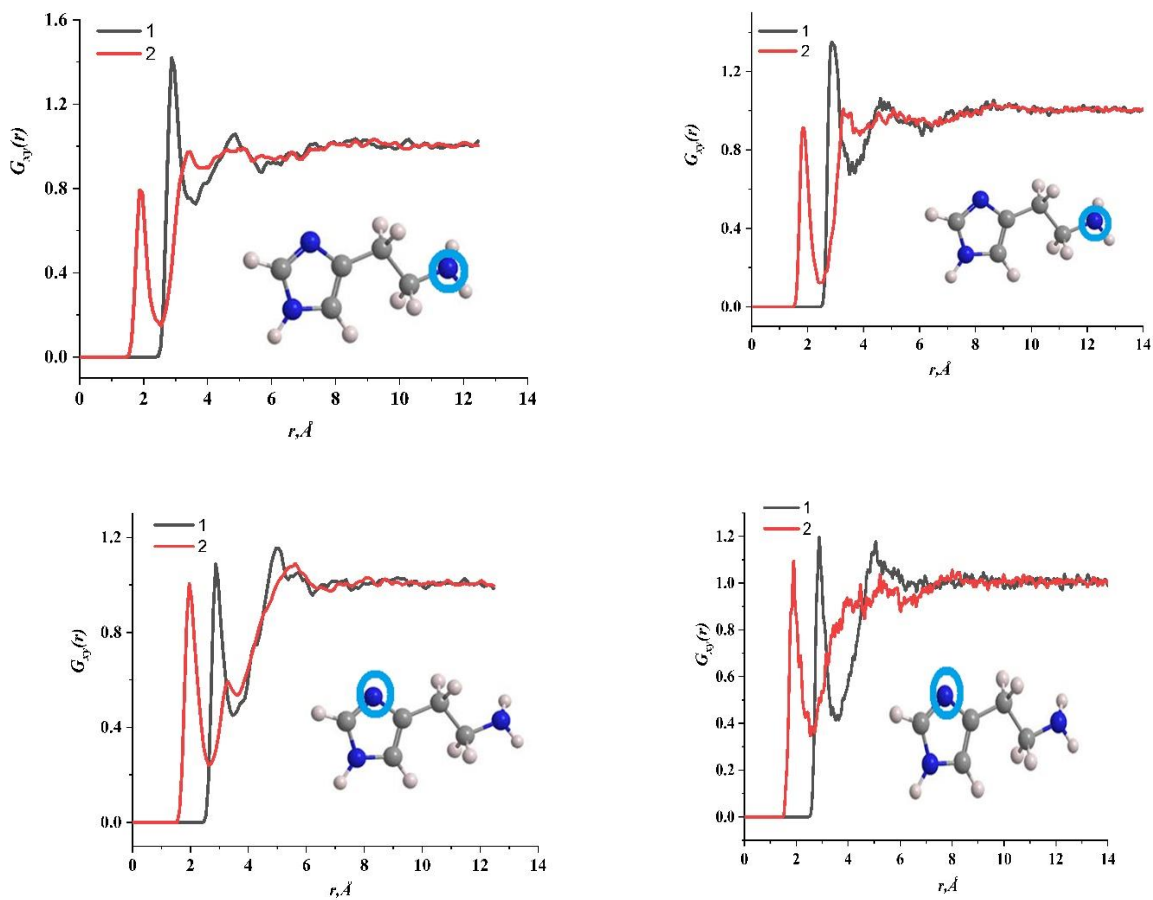

**Fig. 2**. $G_{XY}(r)$ RDF for water-histamine systems at T=300K, in which water is represented within the TIP3P (left column) and SPC/E (right column) models:
upper panel: 1- N45p…$O^W$; 2- N45p…$H^W$
lower panel: 1- N500-3…$O^W$; 2- N500-3…$H^W$

The RDF analysis (Fig. 3) shows that due to the interactions between the C1s, C500-2, C500-4, C500-5, C45 atoms and the atoms of water molecules, they can be located in the vicinity of carbon atoms at a distance from ~3.8 Å (in the case of the C500-2 atom) to ~4.0 Å (in the case of the C1s atom). Moreover, the first hydration shells with a size of ~4.8 Å are formed around the C500-2 and C1s atoms of histamine. The second hydration shell around these atoms reaches ~8.0 Å.

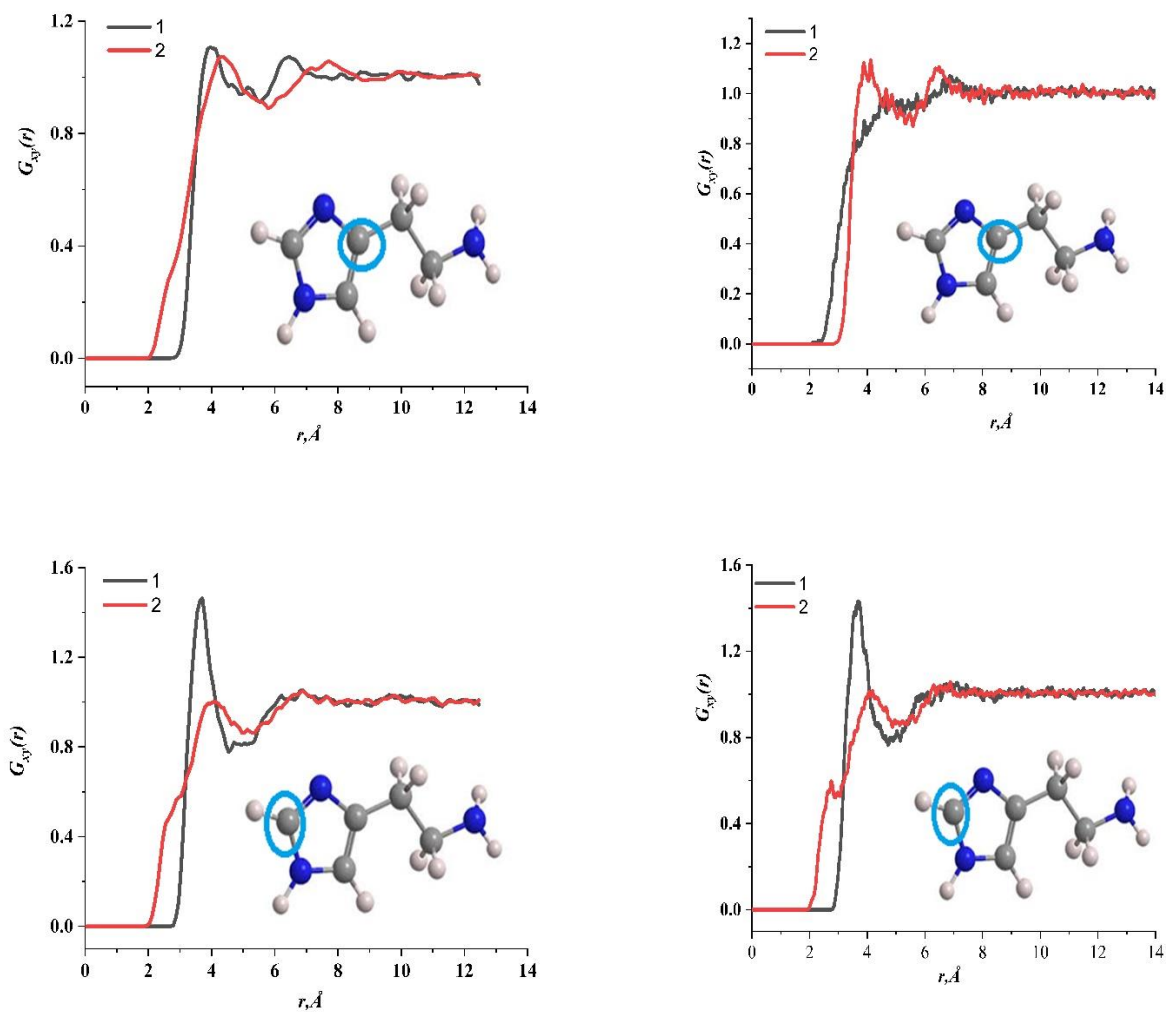

**Fig. 3.** $G_{XY}(r)$ RDF for water-histamine systems at T=300K, in which water is represented within the TIP3P (left column) and SPC/E (right column) models:
upper panel: 1- C1s…$O^W$; 2- C1s…$H^W$
lower panel: 1- C500-2…$O^W$; 2- C500-2…$H^W$

It should be noted that the RDF (Fig.3) shows that the choice of a model representation for describing the interactions between water and histamine molecules does not affect the parameters of the free space between histamine and water atoms and constitutes ~4.0 Å in the vicinity of the carbon atoms of the histamine molecule. RDF (Fig. 4) shows that due to the interaction between the hydrogen atoms of the

NH$_2$ group of histamine and the oxygen atoms of water molecules, HBs with lengths of ~2.0 Å and ~3.2 Å can be formed. In turn, due to the interaction of hydrogen atoms bound to the carbon atoms of the imidazole ring with a relatively low probability (~0.8) lead to the formation of HBs without the formation of clearly defined hydration shells in their vicinity. Due to the interaction between the H45A atom, the CH2 groups can form HBs with a relatively low probability (~0.8), which also affects the underlying processes of forming the local structure of the solution. In addition, RFR (Fig. 4) records the possibility of the formation of a HB with a length of ~ 2.0 Å due to the interaction of a hydrogen atom bound to the nitrogen atom of the imidazole ring with the oxygen atom of the water molecule. Moreover, in the region of this hydrogen atom, a clearly defined first hydration shell (with a size of ~ 2.8 Å) and a second hydration sphere (with a size of ~ 6.5 Å) are observed, which indicates a change in the local density of water.

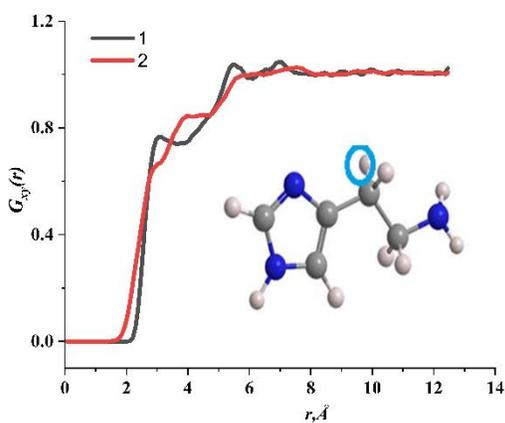
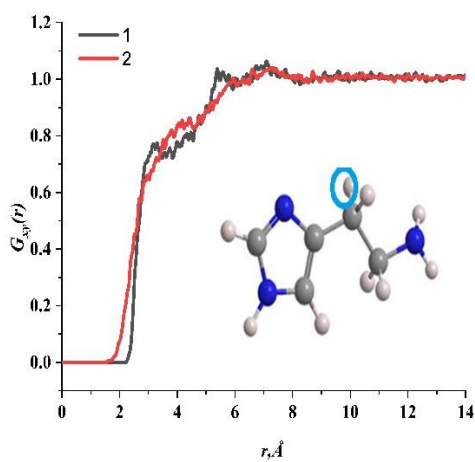

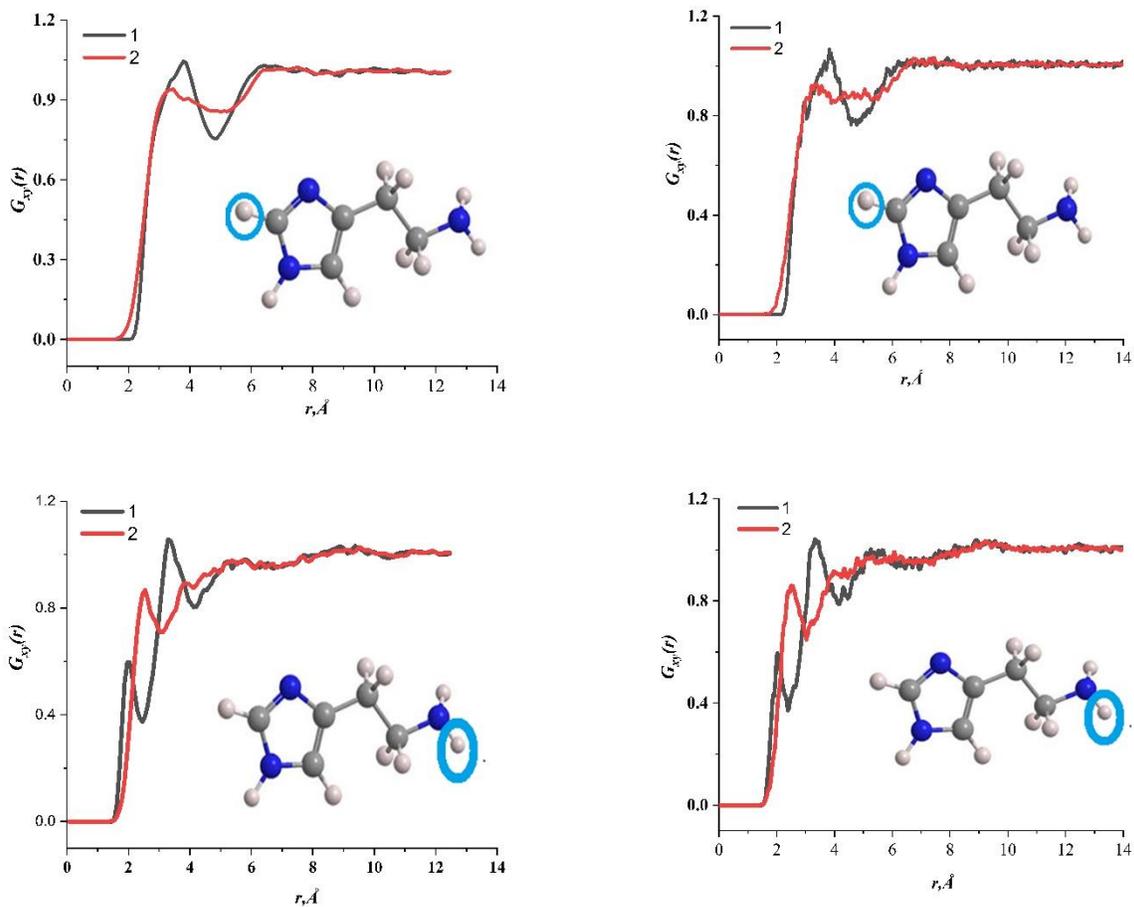

**Fig. 4.** $G_{XY}(r)$ RDF for water-histamine systems at T=300K, in which water is represented within the TIP3P (left column) and SPC/E (right column) models:
upper panel: 1- H1C…$O^W$; 2- H1C…$H^W$
middle panel: 1- H11C…$O^W$; 2- H11C…$H^W$
lower panel: 1- H45N-2…$O^W$; 2- H45N …$H^W$

The analysis of the RDF (Fig. 2-4) shows that the choice of model representation for describing the interactions between water and histamine molecules does not affect the positions of the maxima, probabilities and shape of the RDF [11]. This indicates that the key structural characteristics of the first solvation shells remain stable regardless of the water model. Thus, in the molecular dynamics trajectories of water-histamine interactions, the first solvation shells, determined by the positions of the RDF maxima around the nitrogen atoms, are at comparable distances in both the TIP3P and SPC/E simulations. This is consistent with the general conclusion that different three-point water models under the same thermodynamic conditions reproduce similar positions of the first $G_{XY}(r)$ maxima, although the peak height and structuring outside the first shell may differ slightly due to the difference in the parameterization of the potentials, which is well illustrated by the example of pure water for TIP3P and SPC/E [16]. The RDF (Fig. 2) demonstrate that the distribution of water around the nitrogen atoms of histamine (e.g., N45p, N500-1, N500-3) forms clear first peaks at characteristic distances reflecting the first solvation shell, which correlates with the conclusions of the work [12]. That is, water molecules are located at close distances to the key functional groups of histamine, and the differences between the models are manifested by a slight change in the shape and RDF values for the first maxima, but not in their position, which correlates with the conclusions of the work [13]. Such a stable nature of the RDF reflects the fundamental physicochemical principles of solvation: the local density of water around a dissolved organic molecule is determined primarily by its geometry and intermolecular distances in the liquid, and not by the subtleties of a specific parameterization of the three-point water model [14]. Taken together, the data obtained indicate that the choice of model representation (TIP3P or SPC/E) for describing water interactions with histamine has a negligible effect on the positions of the maxima, probabilities, and shape of the RDF, which is important when comparing the structural characteristics of different force fields in molecular dynamics.

*Dynamics of the water-histamine system*

It was previously established that the use of TIP3P in modeling the properties of solutions can lead to overestimated diffusion coefficients in the systems under study and underestimated lifetimes of solute-water hydrogen bonds [15]. Qualitatively studying the effect of the choice of model representation for water on the dynamics of the components of the histamine-water solution at infinite dilution at T=300K allows the analysis of the mean square displacement (MSD) function. This function describes the displacement of the center of mass of the *i*-th molecule over the time interval from *0* to *t* and is determined by the following relation [16]:

$$\langle r^2(t) \rangle = \frac{1}{N} \langle \sum_{i=1}^{N} | \Delta r_i(t) |^2 \rangle. \tag{4}$$

Study of the values of the parameter $\alpha$ (Fig. 5) in the logarithmic dependence

$$\log \langle r^2(t) \rangle \sim a \log(t) \tag{5}$$

makes it possible to qualitatively distinguish time intervals in which the nature of the motion of the system components changes. Such changes are associated with the restructuring of the local structure of the solution, changes in diffusion mechanisms. In addition, given that the histamine-water system can be considered as a mixture of large (histamine) and small (water) particles with a mass ratio $\alpha = \frac{M_{histamine}}{M_{water}} > 1$ and according to the conclusions of [18], the structural rearrangement in the histamine-water solution is determined by the processes of short-term relaxation and diffusion on short times. Therefore, it is advisable to choose a short time scale (t < 1000 ps) for analysis when studying the dynamics of the solution components and the influence of the choice of the water model on the dynamics of the studied system.

Fig. 5 presents the functions $\langle r^2(t) \rangle$ for water (top panel) described within the SPC/E model (Fig. 5 (1)) and TIP3P (Fig. 5 (2)). At times less than 200ps, the values of $\langle r^2(t) \rangle$ water for different models coincide. At times greater than 200ps, the values of the function $\langle r^2(t) \rangle$ for water described within the SPC/e model exceed the values of the function $\langle r^2(t) \rangle$ for water described within the TIP3P model. However, the behavior of the $\log \langle r^2(t) \rangle$ function for both SPC/E water and TIP3P water models record the coincidence of water diffusion mechanisms at the same time intervals.

Namely, at small times, water moves in a slow sub-diffuse mode, which passes into a practically diffuse mode at relatively large time intervals. In turn, the behavior of the function $\langle r^2(t) \rangle$ for a histamine molecule in water under different model representations correlates with the behavior of the corresponding $\langle r^2(t) \rangle$ water functions. Namely, the function $\langle r^2(t) \rangle$ for a histamine molecule in water, which is described within the TIP3P framework, grows much faster than $\langle r^2(t) \rangle$ of histamine in water, which is modeled using the SPC/E model. At time intervals of less than 200 ps, the mechanisms of histamine movement in the studied systems can be described within the framework of slow sub-diffuse. At relatively large time intervals in the histamine-water system (TIP3P model), the movement of the histamine molecule changes from slow sub-diffuse to diffuse. No changes in the nature of histamine diffusion in the histamine-water system (SPC/E model) are observed. The obtained result correlates with the conclusions [17,18], which state that the use of the TIP3P model representation for the description of water leads to larger values of the ligand diffusion coefficients in water, while the use of the SPC/E model for water allows for a better description of the solvation processes.

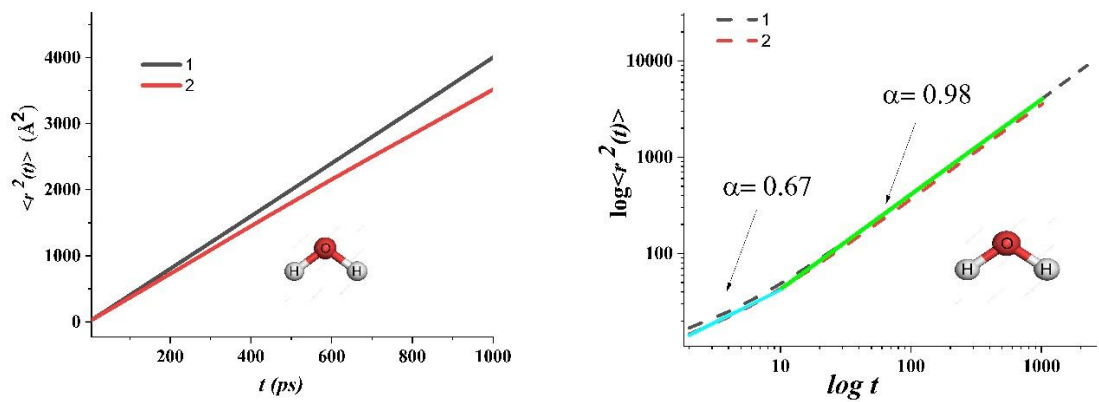

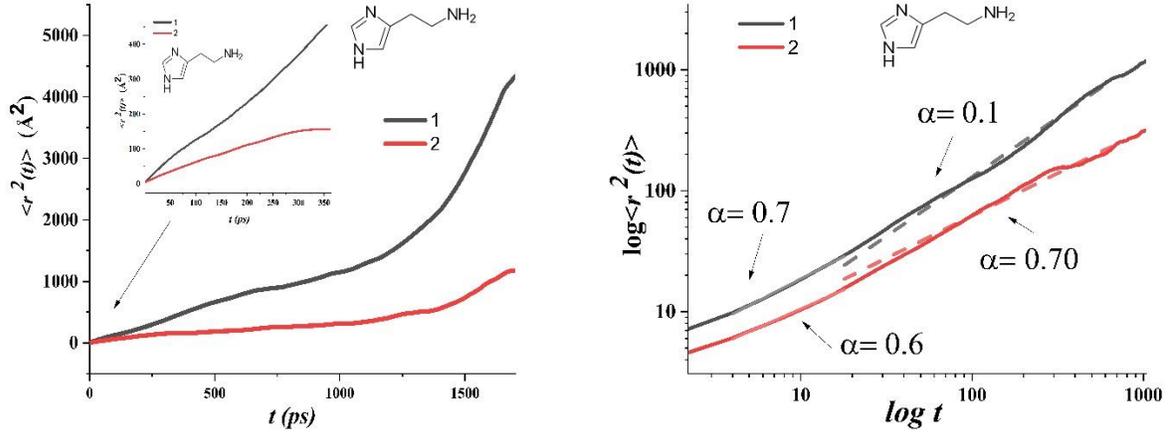

**Fig. 5** Time dependence of the function $\langle r^2(t) \rangle$ (a), the function $log\langle r^2(t)\rangle$ (b) of the center of mass of a water molecule (upper panel) and a histamine molecule (lower panel) in the histamine-water system (1- SPC/E, 2- TIP3P water models) at T=300K

This is confirmed by the values of the self-diffusion coefficients of the components of the studied systems obtained as a result of calculations, which can be obtained using the Green-Kubo relation, based on the integration of the autocorrelation velocity function (VAF) [10], which is a function of the velocity of a moving tracker particle (water or histamine) $\vec{V}(t) = \vec{p}/m$ through the liquid in the x direction and is defined as:

$$Z(t) = \frac{1}{3}\langle \vec{V}(0)\vec{V}(t) \rangle = \langle V_x(0)V_x(t) \rangle, \qquad (6)$$

where $\vec{V}(0)$ is the component of the particle velocity along its initial direction of motion, averaged over the initial conditions. The diffusion coefficient of the components of the water-histamine systems in the studied systems was calculated according to the Green-Kubo relation:

$$D = \frac{1}{3}\int_0^\infty \langle V(t)V(0) \rangle dt. \qquad (7)$$

The calculated value of the water diffusion coefficient in the histamine-water system (SPC/E model) is $2.85*10^{-9}$, m²/s, which correlates relatively well with the experimental value of the water diffusion coefficient [19]. The value of the histamine diffusion coefficient in the histamine-water system (SPC/E model) is $0.38*10^{-9}$, m2/s. In turn, the value of the water diffusion coefficient in the histamine-water system (TIP3P model) is $5.94*10^{-9}$, m²/s, which significantly exceeds the experimental values of the water diffusion coefficient. The value of the histamine diffusion coefficient in this system is $4.61*10^{-9}$, m²/s. It should be noted that when studying the dynamic behavior in the "water–histamine" system at T= 300K, it can be assumed that water at T < 350K [20,21] can be represented as a practically harmonic system with relatively rare "jumps" of water molecules from one equilibrium position to another [22]. At the same time, according to Frenkel [23], the diffusion of components of highly diluted aqueous solutions can be described within the framework of the vibrational-jump diffusion model. Within the framework of such a model approach, a particle in an equilibrium position can perform one or two oscillations. The complexity of the description lies in the lack of a clear criterion for separating the purely vibrational and vibrational-jump mechanisms of diffusion of a particle (water or solute) in the case when the particle makes only one or two oscillations between the jumps. This problem can be solved by analyzing the time dependence of the derivative of the correlation scattering functions) $F_s(q,t)$ [24,25], which allows us to determine the values of the corresponding times at which the diffusion mechanisms change in the system. The scattering function $F_s(q,t)$ can be represented as:

$$F_s(q,t) = \frac{1}{N} \sum_{j=1}^{N} \exp[i\vec{q} \cdot (\vec{r}_j(t) - \vec{r}_j(0))], \tag{8}$$

where $\vec{r}_j$ — is the position vector of the *j*-th particle.

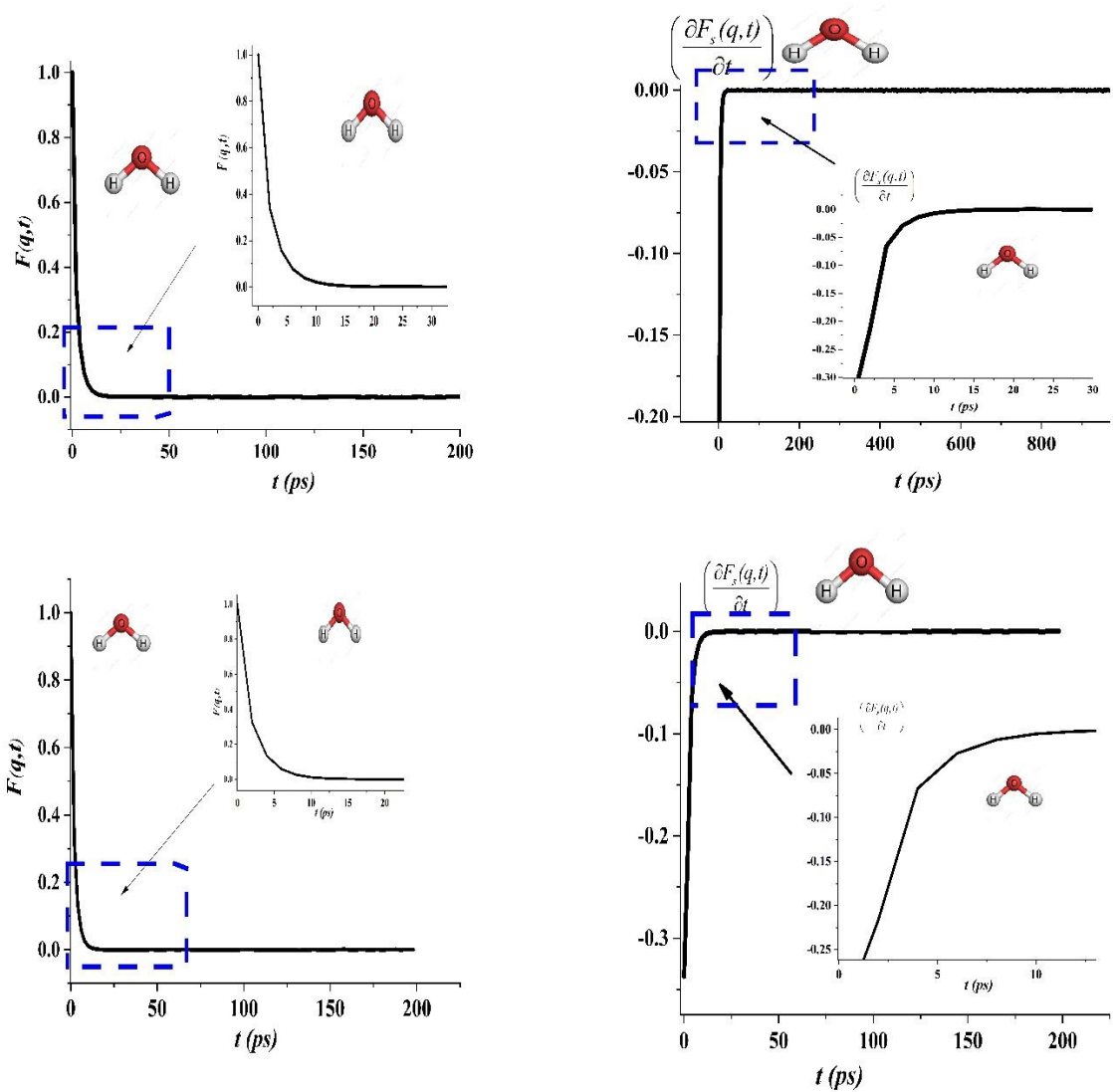

**Fig. 6.** Time dependence of the function $F_s(q,t)$ and $\left(\dfrac{\partial F_s(q,t)}{\partial t}\right)$ the center of mass of water on time in the histamine-water system for different water models at T=300K: TIP3P – upper panel, SPC/E – lower panel

The first inflection point of the graph $\left(\dfrac{\partial F_s(q,t)}{\partial t}\right)$ for water in both the TIP3P and SPC/E water models is reached at approximately ~5 ps. The second inflection point of the function $\left(\dfrac{\partial F_s(q,t)}{\partial t}\right)$ is fixed at approximately ~8 ps for water.

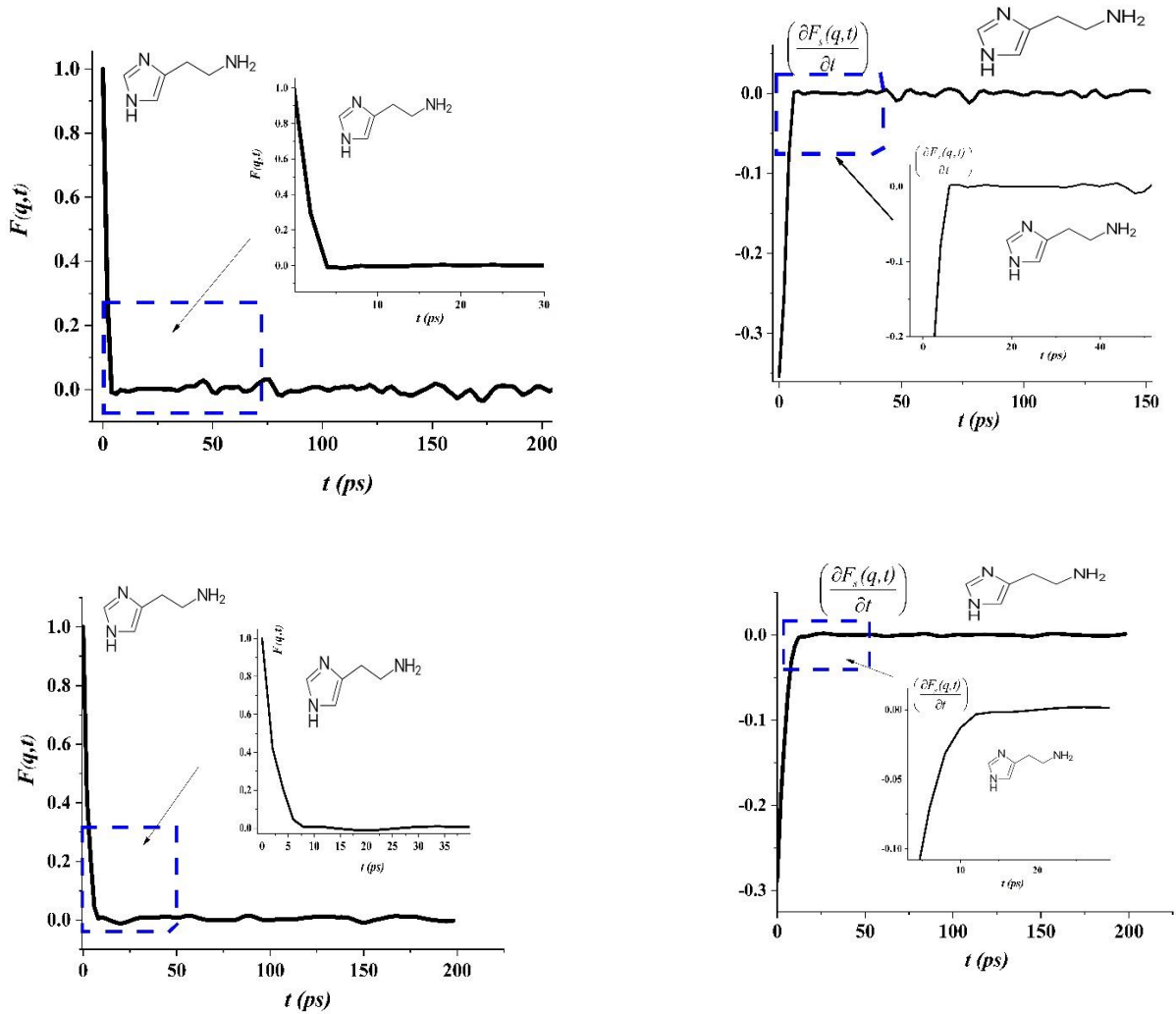

**Fig. 7**. Time dependence of the function $F_s(q,t)$ and $\left(\dfrac{\partial F_s(q,t)}{\partial t}\right)$ center of mass of histamine on time in the histamine-water system at T=300K:
TIP3P water model – upper panel,
SPC/E water model – lower panel

The first time interval to the inflection point $\left(\dfrac{\partial F_s(q,t)}{\partial t}\right)$ for histamine is fixed at ~10 ps. The second inflection point of the graph $\left(\dfrac{\partial F_s(q,t)}{\partial t}\right)$ for histamine is at ~15 ps. At times less than ~3 ps, the functions $\left(\dfrac{\partial F_s(q,t)}{\partial t}\right)$ for water and histamine coincide. This indicates that at small time intervals less than ~3 ps the mechanisms of motion of histamine and water are the same. The obtained result indicates the independence of the time intervals of changes in diffusion mechanisms from the choice of the model representation of water, and is determined by the physical properties of the studied histamine-water system.

**CONCLUSIONS**

In the work, the structural characteristics of histamine hydration in a dilute aqueous solution at a temperature of T = 300 K were studied using the molecular dynamics method using two common three-point water models – TIP3P and SPC/E. Analysis of radial distribution functions allowed us to characterize in detail the spatial organization of water molecules around the key atoms of histamine, in particular the nitrogen atoms of the imidazole ring and the primary amino group, as well as the carbon atoms of the molecule. The work shows that histamine hydration has a clearly pronounced shell character. Stable first hydration shells with characteristic radii of ~2.7 Å – ~3.0 Å are formed around the nitrogen atoms. The existence of a second and, in some cases, several hydration shells was established for the atoms of the imidazole ring, which indicates a complex solvation structure. Around the carbon atoms of histamine, the first hydration shells are more diffuse and localized at larger distances (~3.8 Å – ~ 4.8 Å), which corresponds to a weaker specific interaction with water. The key result of the work is the establishment that the choice of the water model (TIP3P or SPC/E) has practically no effect on the positions of the first RFR maxima, their radii, and the general topology of the first solvation shells of histamine. This indicates that the main structural characteristics of hydration are determined by the geometry and chemical nature of the histamine molecule, and not by the details of the parameterization of a specific three-point water model. The results obtained demonstrate the structural stability of the histamine hydration shell with respect to the choice of the water model and confirm the correctness of using both TIP3P and SPC/E for the analysis of the structural properties of systems of the type "imidazole ligand – water". The dynamics of the system is described within the framework of the oscillatory-jump diffusion model: on small time intervals, the motion of water and histamine has a common sub-diffusive oscillatory character. It is shown that the transition to a practically diffusive mode of motion on relatively large time intervals is recorded only in the case of using the TIP3P water model, while in the system simulated using SPC/E, the motion of the components remains subdiffusive throughout the entire studied time range. Such a change in the nature of diffusion when using the TIP3P water model may be the reason for obtaining overestimated values of diffusion coefficients when modeling aqueous solutions of ligands. The SPC/E model, on the contrary, provides stronger solvation and a physically correct description of diffusion mobility, consistent with experimental data.

**Acknowledgments**
N. Atamas acknowledges the support of the BOKU University (Vienna, Austria) financed through the MSCA4Ukraine project, funded by the European Union.